# *In vivo* high resolution human retinal imaging with wavefront correctionless full-field OCT

PENG XIAO[1], VIACHESLAV MAZLIN[1], KATE GRIEVE[2], JOSE-ALAIN SAHEL[2], MATHIAS FINK[1], AND A. CLAUDE BOCCARA[1,*]

[1]*Institut Langevin, ESPCI Paris, PSL Research University, 1 rue Jussieu, 75005 Paris, France*
[2]*CHNO des Quinze Vingts, Institut de la Vision, INSERM-DHOS CIC 503, Sorbonne Universités, UPMC Univ Paris 06, CNRS, 28 rue de Charenton, 75012 Pairs, France*
*\*Corresponding author: claude.boccara@espci.fr*

**As the lateral resolution of FFOCT with spatially incoherent illumination has been shown to be insensitive to aberrations, we demonstrate high resolution *en face* full-field OCT (FFOCT) retinal imaging without wavefront correction in the human eye *in vivo* for the first time. A combination of FFOCT with spectral-domain OCT (SDOCT) is applied for real-time matching of the optical path lengths (OPL) of FFOCT. Through the real-time cross-sectional SDOCT images, the OPL of the FFOCT reference arm is matched with different retinal layers in the FFOCT sample arm. Thus, diffraction limited FFOCT images of multiple retinal layers are acquired at both the near periphery and the fovea. The *en face* FFOCT retinal images reveal information about various structures such as the nerve fiber orientation, the blood vessel distribution, and the photoreceptor mosaic.**

During its 25 years of development, optical coherence tomography (OCT) has become a powerful imaging modality in biomedical and clinical studies [1,2]. It has achieved great success in ophthalmology, especially in retinal imaging [3]. OCT imaging has revolutionized the diagnosis and treatment of many retina diseases, for which it has been entitled the "virtual biopsy" for human retina. Compared with typical retinal imaging modalities like fundus camera [4] or scanning laser ophthalmoscopy (SLO) [5], in which axial resolution is limited by the finite size of the eye's pupil (i.e. the numerical aperture (NA) of the eye), OCT offers much higher axial sectioning as the lateral and axial resolutions are decoupled. The high-resolution cross-sectional depth exploration of the retinal layers offers important information about pathologies for early diagnosis of disease and for tracing disease evolution [6-8]. Nevertheless, due to the conventional use of imaging devices like fundus camera or SLO, ophthalmologists often ask for *en face* images of OCT. Thanks to the speed improvement of OCT systems, *en face* retinal images can be obtained by real-time 3D imaging [9-11]. Nevertheless, due to the requirement of large depth of focus (typically the full retinal thickness), low numerical aperture (NA) is typically used in traditional OCT; resulting in relatively low spatial resolution compared with high NA systems. In order to be able to realize close to diffraction-limited lateral resolution in OCT retinal imaging, complex hardware adaptive optics (AO) [12-14] or computational AO [15,16] would also be needed to correct the aberrations induced by the imperfections of the cornea and lens in the anterior chamber.

Full-field OCT (FFOCT) is a kind of parallel OCT that takes *en face* images perpendicular to the optical axis without scanning. By using high NA microscope objectives in a Linnik interferometer, FFOCT is able to achieve standard microscope spatial resolution [17]. With spatially incoherent illumination, cross-talk is severely inhibited in FFOCT compared with wide-field OCTs that use spatially coherent illumination [18]. Moreover, the use of spatially incoherent illumination in FFOCT offers another advantage that we have demonstrated recently: the lateral resolution is independent of aberrations that only affect the FFOCT signal level (or signal to noise ratio (SNR)) [19], which is not the case for OCTs with spatially coherent illumination. Thus in terms of human retinal imaging, FFOCT can keep the near diffraction-limited lateral resolution. Since FFOCT detects the amplitude of the interference field, the signal reduction induced by the eye aberrations is proportional to the square root of the Strehl ratio, meaning that only aberrations with large values affect the FFOCT image quality. These properties make FFOCT is a promising imaging modality for high resolution *en face* retinal imaging.

To successfully apply FFOCT to *in vivo* human retinal imaging, the obstacles that needed to be resolved include optical path length (OPL) matching difficulties and eye motion. In this letter, we propose to combine an FFOCT system with a spectral-domain OCT (SDOCT) system to realize real-time matching of the OPL for FFOCT through the cross-sectional images of SDOCT. The OPL of the FFOCT reference arm is matched with various retinal layers in the FFOCT sample arm by translating the whole system along the optical axis for *en face* FFOCT retinal imaging. By implementing a new high speed camera working at up to 750 Hz, eye motion is sufficiently reduced during the image acquisition to record the interferometric signal. With the combined system, we have achieved *in vivo* cellular FFOCT human retina imaging for the first time to the best of our knowledge without applying wavefront correction [20]. FFOCT retinal images of the near periphery and the fovea have been acquired and compared with images acquired with the only currently commercially available clinical AO retinal camera.

The system schematic is showed in Fig. 1.: A customized FFOCT system is combined with the Thorlabs GANYMEDE-II SDOCT system. In the FFOCT part, an LED with λ=850 nm center wavelength and 30 nm bandwidth (Thorlabs) is used as the incoherent light source, giving an axial resolution of 8 μm in water. The illumination beam is split into the sample arm and reference arm with a 50:50 cubic beamsplitter (BS). An Olympus 10 X/0.25 NA Plan Achromat objective is used in the reference arm with a silicon mirror supported by a piezoelectric transducer (PZT) and placed at the focal plane of the objective. The human eye is aligned in the sample arm along the optical axis with the temples and chin pressed against the headrest in order to minimize head movements, and a fixation target is used to reduce lateral eye motion. Dispersion is balanced with a glass window between the microscope objective in the reference arm and the human eye. The back-reflected beams from the two arms are recombined with the BS and imaged onto a high-speed (750 fps) CMOS camera (Q-2A750-Hm/CXP-6, ADIMEC) for FFOCT imaging. The camera sensitivity is calculated to be 77 dB for a single image with a speed of 750 Hz for 2 Mpixels. The Thorlabs GANYMEDE-II SDOCT system uses a broadband SLD with center wavelength of 930 nm, giving an axial resolution in water of 4.5 μm. It comes with a scanning system kit that contains X and Y scanners while the reference arm is customizable. The SDOCT has a sensitivity of about 96 dB at the highest A-scan rates of 36 kHz with 1024 axial pixels.

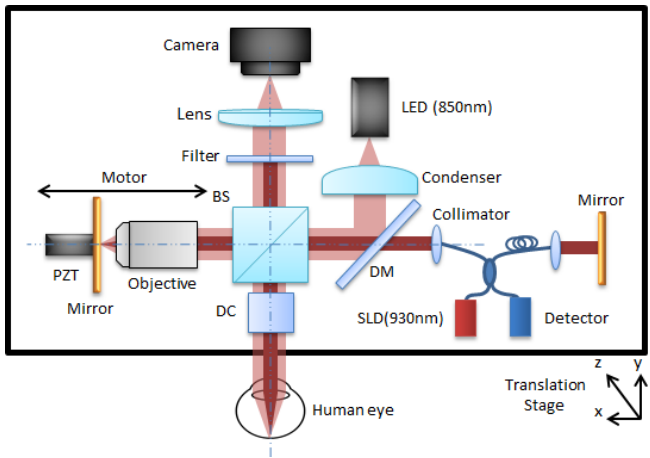

Fig. 1. Schematic of the combined system of FFOCT with SDOCT. BS: beamsplitter; DC: dispersion compensation; DM: dichroic mirror; PZT: piezoelectric transducer.

As illustrated in Fig. 1, the two OCTs are combined by joining the sample arm of the SDOCT system with the illumination path of the FFOCT through a dichroic mirror (DM). Both systems can work independently yet simultaneously. The combined system is mounted on a platform with 3-dimensional mobility offered by precise translating motors. The OPL of the SDOCT customized reference arm is matched with the OPL of the SDOCT sample arm to the silicon mirror in the FFOCT reference arm. In this way, we are able to take advantage of the real-time imaging ability and the larger depth of field of the SDOCT line-scanning cross-sectional image for real-time matching of the OPL of the FFOCT. This is relatively difficult to balance within 1 μm for the two arms with different geometries, but is achieved by overlapping the SDOCT image of the FFOCT reference mirror and the various retinal layers and translating the whole system with respect to the eye (Visualization 1). By simultaneously recording with both FFOCT and SDOCT, the SDOCT cross-sectional image indicates the depth of the *en face* FFOCT slice. The combined system results in an FFOCT retinal imaging field of view of 2.4° × 2.4° (720 μm × 720 μm) with an illuminating power of 1.3 mW and an SDOCT retinal line-scanning range of 1.6° (480 μm) with a power of 250 μW. Powers applied to the retina are below the maximum permissible exposure according to ANSI [21] and ISO [22] standards.

For the *in vivo* experiments we conducted in this paper, the FFOCT camera is working at a speed of 400Hz. With 2-phase modulation, the FFOCT images can be recorded at a frequency of 200Hz, which is fast enough to freeze the eye motion during an image acquisition according to the studies demonstrated in [23,24]. An image stack of 40 FFOCT retinal images is acquired for one imaging region at a specific depth during a period of 200 ms. The effects of the lateral eye movements on one image stack are then corrected with the ImageJ plugin "Template Matching" [25] for lateral motion correction through cross correlation of the detected structural signals and several images are averaged to improve the SNR.

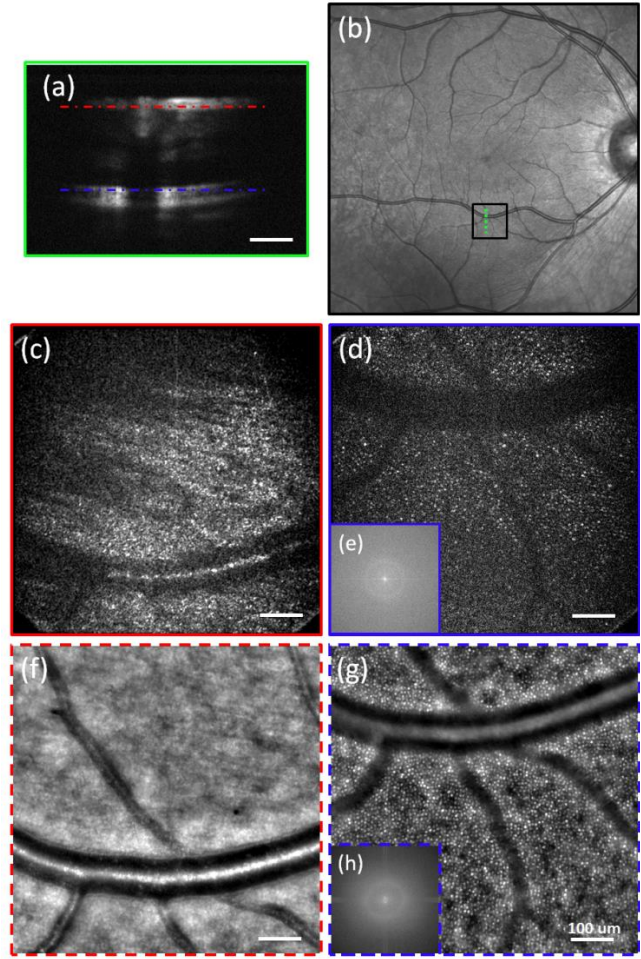

Fig. 2. *In vivo* human retinal imaging of near periphery at 6° eccentricity inferior to the foveal center with a field of view of 2.4°×2.4°. (a) The SDOCT cross-sectional image of the imaging position with the red (RNFL) and blue (IS/OS) dashed lines indicating the FFOCT imaging depth. (b) Fundus photography with the black box indicating the FFOCT imaging area and the green dashed line showing the SDOCT scanning position. (c-d) *In vivo* FFOCT image of at 6° inferior to the fovea, at the RNFL (c) and IS/OS photoreceptor layers (d) without AO. (e) The 2D power spectra of (d) showing the Yellot's ring, the radius of which is related to the cone photoreceptor spacing. (f-h) The AO retinal camera image around the RNFL (f) and IS/OS photoreceptor layer (g) at the same retinal location and the 2D power spectra (h) of (g). Scale bar: 100 μm.

Before conducting the *in vivo* experiments, informed consent was obtained for the subject and the experimental procedures adhered to the tenets of the Declaration of Helsinki.

The first experiment on the *in vivo* human retina was located in the near periphery. The SDOCT begins real-time line scanning of the retina. By looking at the fixation target and changing the fixation target position, retinal imaging was performed at about 6° eccentricity inferior to the foveal center. Note that by minimizing the light level in the room the subject had a pupil of 4.5 mm during the experiment, and no pupil dilation was applied. The expected FFOCT diffraction limited lateral resolution would be 4 μm. By translating the system along the optical axis, the SDOCT image of the FFOCT reference mirror is overlapped with the SDOCT retinal image of the retinal nerve fiber layer (RNFL) and the inner/outer segment (IS/OS) photoreceptor layer. In the meantime, the FFOCT imaging is launched at each position with a stack of 40 images, which are processed by ImageJ for lateral motion correction and averaged. Experimental results are shown in Fig. 2. Fig. 2 (a) shows the cross-sectional SDOCT image of the retinal layers with the red (RNFL) and blue (IS/OS) dashed line indicating the overlapped positions with the SDOCT image of FFOCT reference mirror for *en face* FFOCT imaging. Fig. 2 (b) shows the fundus photography (acquired with SPECTRALIS retinal imaging platform, Heidelberg Engineering, Germany) with the black box indicating the FFOCT imaging region and the green dashed line indicating the SDOCT scanning location. Fig.2 (c) is the averaged FFOCT image of the RNFL, in which the orientation of the nerve fibers are visible as well as a large blood vessel. Fig.2 (d) is the FFOCT image of the IS/OS photoreceptor layer, which clearly shows the cone photoreceptor mosaic as well as the shadows of the distribution of blood vessels. The cone photoreceptors have an averaged diameter of about 5.5 - 6 μm, corresponding to a real cone photoreceptor size of around 4 μm in diameter with our lateral diffraction limited resolution of about 4 μm. The power spectra obtained by Fourier transform of the FFOCT image of the IS/OS layer shows the Yellot's ring (Fig.2 (e)), which corresponds to the spatial frequency of the cone photoreceptors [14,26,27], indicating that we are resolving the cone photoreceptors in this FFOCT image. The circumferential-averaged power spectrum gives a maximum at 31.6 cyc/degree, corresponding to a cone spacing of about 9.5 μm. This matches with the values given in references [14,26] for cone spacing at around $6° - 7°$ eccentricity. Retinal imaging has also been done using a commercial AO retinal camera (RTX1™, Imagine Eyes, France) around the same retinal layers of the same eye at the same location for comparison. The images are shown in Fig. 2 (f,g). The 2D power spectrum (Fig. 2 (h)) of the photoreceptor image shows the Yellot's ring with the circumferential-averaged power spectrum maximum at 31.9 cyc/degree, corresponding to a cone spacing of about 9.4 μm. This is close to what we obtained with the FFOCT image. The FFOCT RNFL image (Fig. 2(c)) shows advantages compared with the AO retinal camera RNFL image (Fig. 2 (f)) as the axial sectioning ability is poorer in the AO retinal camera, making it difficult to select only specific retinal layers. Thus the small blood vessels, which are supposed to be under the RNFL appear in the AO retinal camera image. For the IS/OS photoreceptor layer image, the SNR of the FFOCT image is lower compared with the AO retinal camera image (Fig. 2 (g)). This might be due to the fact that we are not correcting the aberrations in the eye, which affects the signal level of FFOCT as explained in reference [19]. Also, the smaller axial sectioning offered by FFOCT selects the signal from a thinner retinal layer, which would give a relatively lower signal level.

Imaging was also performed in the fovea to image the IS/OS photoreceptor layer. By fixating on the scan line of the SDOCT, the image is captured at the foveal center. The SDOCT image of the fovea is shown in Fig. 3 (a), which shows the IS/OS layer and RPE layer of the retina. By translating the system, the SDOCT image of IS/OS layer is overlapped with the SDOCT image of the FFOCT reference mirror (indicated by the red dashed line in Fig. 3 (a)) and 40 FFOCT images are taken, registered, and averaged. Again, the black box in the fundus photograph shown in Fig. 3 (b) indicates the FFOCT imaging region, and the green dashed line indicating the SDOCT scanning location. Processed by ImageJ, the final FFOCT image of the fovea is shown in Fig. 3 (c) with zoomed in areas at about 1° eccentricity (Fig. 3 (d)) and the fovea center (Fig. 3 (e)). As the lateral resolution is not sufficient to resolve all of the cones in the fovea center, the 2D power spectrum (Fig. 3 (f)) of the FFOCT image shows no Yellot's ring. Nevertheless, some cones with a higher signal are detectable and appear in the image. These structures are not speckles as they are the basis for the cross correlation to correct the lateral motion artefacts before averaging. The imaging location of this experiment is also further confirmed by imaging the fovea while focused at the photoreceptor layer with the AO retinal camera. These images are shown in Fig. 3(g-i) for comparison. Note that even with AO, the AO retinal camera is also not able to resolve the cone photoreceptors at the foveal center as no Yellot's ring appears in the 2D power spectrum (Fig. 3 (j)).

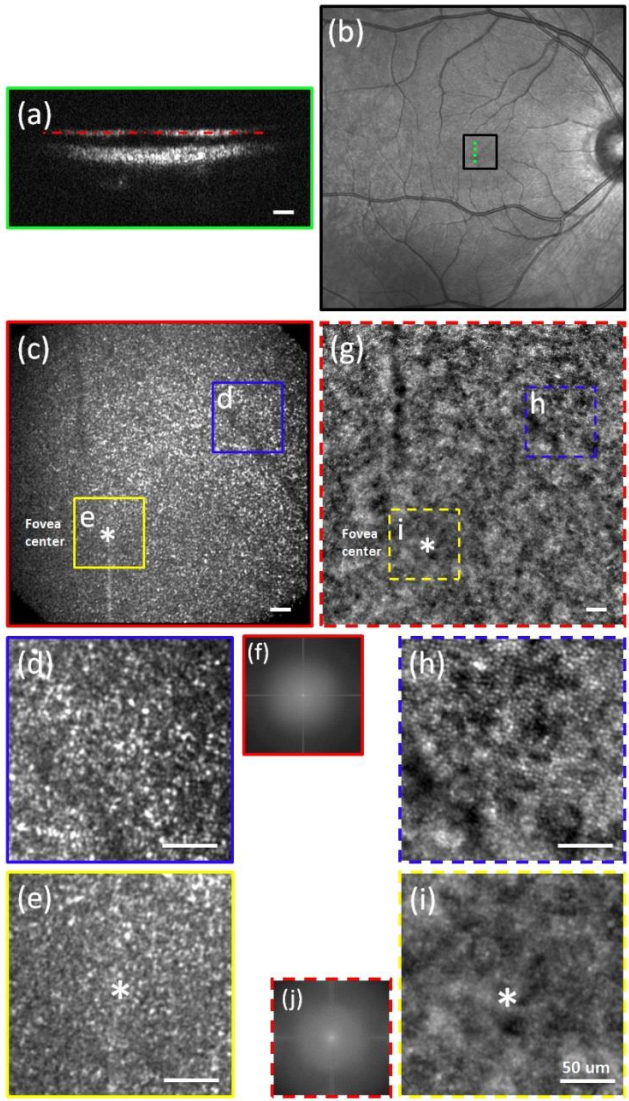

Fig. 3. *In vivo* human retinal imaging at the fovea. (a) The SDOCT cross-sectional image of the foveal center where the red dashed line indicates the depth at which the FFOCT image is taken. (b) Fundus photograph with the black box indicating the FFOCT imaging area and the green dashed line showing the SDOCT scanning position. (c) *In vivo* FFOCT image of the human fovea without AO and the zoomed in areas at about 1° eccentricity (d) and the fovea center (e). (f) The 2D power spectrum of (c). (g-j) The AO-fundus camera image of the same locations as (c-e) for comparison. (j) The 2D power spectrum of (g) Scale bar: 50 μm.

In conclusion, we have resolved the OPL matching difficulties of FFOCT for *in vivo* human retinal imaging by combining it with an SDOCT system. Real-time OPL matching is possible with the cross-sectional SDOCT images. Eye movements are frozen during FFOCT image acquistion by implementing a high speed CMOS camera working at up to 750Hz. Thanks to the spatial resolution merit of FFOCT with spatially incoherent illumination, high resolution *en face* FFOCT cellular retinal imaging has been achieved in humans *in vivo* without AO or wavefront post processing, for the first time to the best of our knowledge, both in the near periphery and the fovea at different retinal layers showing structural information such as nerve fiber orientation, blood vessel distribution as well as the cone photoreceptor mosaic. Our current system has insufficient SNR to image other retinal layers such as the ganglion cell and retinal pigment epithelium as we are not applying adaptive optics. As we have demonstrated in [28], a compact implementation of a transmissive wavefront corrector without strict pupil conjugation has been utilized for low order aberration correction to improve the FFOCT signal level. Thus for human eyes, in which low order aberrations dominate, an adaptive liquid lens [29] could be implemented to correct only large defocus and astigmatism to improve the SNR.

**Funding.** This work was supported by the HELMHOLTZ synergy funded by the European Research Council (ERC grant agreement #610110).